\documentclass[aps,prl,twocolumn,floatfix]{revtex4}
\usepackage{graphicx}
\usepackage{bm}

\begin{document}
\title{Fractional vortices and composite domain walls in flat nanomagnets}
\author{Oleg Tchernyshyov}
\author{Gia-Wei Chern}
\affiliation{Department of Physics and Astronomy, 
The Johns Hopkins University, Baltimore, Maryland, 21218, USA}
\date{11 September 2005}

\begin{abstract} 
We provide a simple explanation of complex magnetic patterns observed
in ferromagnetic nanostructures.  To this end we identify elementary
topological defects in the field of magnetization: ordinary vortices
in the bulk and vortices with half-integer winding numbers confined to
the edge.  Domain walls found in experiments and numerical simulations
in strips and rings are composite objects containing two or more of
the elementary defects.
\end{abstract}

\maketitle

Topological defects \cite{Chaikin,Mermin79} greatly influence the
properties of materials by catalyzing or inhibiting the switching
between differently ordered states.  In ferromagnetic nanoparticles of
various shapes (e.g. strips \cite{Atkinson03} and rings \cite{Zhu00}),
the switching process involves creation, propagation, and annihilation
of domain walls with complex internal structure \cite{Klaeui03R}.
Here we show that these domain walls are composite objects made of two
or more elementary defects: vortices with integer winding numbers ($n
= \pm 1$) and edge defects with fractional winding numbers ($n = \pm
1/2$).  The simplest domain walls are composed of two edge defects
with opposite winding numbers.  Creation and annihilation of the
defects is constrained by conservation of a topological charge.  This
framework provides a basic understanding of the complex switching
processes observed in ferromagnetic nanoparticles.

In ferromagnets the competition between exchange and magnetic dipolar
energies creates nonuniform patterns of magnetization in the ground
state.  Whereas the exchange energy favors a state with uniform
magnetization, magnetic dipolar interactions align the vector of
magnetization with the surface.  In a large magnet a compromise is
reached by the formation of uniformly magnetized domains separated by
domain walls.  In a nanomagnet magnetization varies continuously
forming intricate yet highly reproducible textures, which include
domain walls and vortices \cite{Klaeui03R,Shinjo00,Castano03}.
Numerical simulations \cite{Gadbois95,McMichael97} reveal a rich
internal structure and complex dynamics of these objects.  For
example, collisions of two domain walls can have drastically different
outcomes: complete annihilation or formation of other stable
textures \cite{Castano03}.  These puzzling phenomena call for a
theoretical explanation.

An elementary picture of topological defects in nanomagnets with a
planar geometry is presented in this Letter.  It is suggested that the
{\em elementary} defects are vortices with integer winding numbers and
edge defects with half-integer winding numbers.  All of the intricate
textures, including the domain walls, are {\em composite} objects made
of two or more elementary defects.

For simplicity, we consider a ferromagnet without intrinsic
anisotropy, which is a good approximation for permalloy.  The magnetic
energy consists of two parts: the exchange contribution $A \int
|\nabla \hat\mathbf{m}|\, d^{3}r$, where $\hat\mathbf{m} =
\mathbf{M}/|\mathbf{M}|$ is the unit vector pointing in the direction
of magnetization $\mathbf{M}$, and the magnetostatic energy
$(\mu_0/2)\int |\mathbf{H}|^2 \, d^3{r}$.  The magnetic field
$\mathbf{H}$ is related to the magnetization through Maxwell's
equations, ${\bm \nabla} \times \mathbf{H} = 0$ and ${\bm \nabla}
\cdot (\mathbf{H} + \mathbf{M}) = 0$.  Apart from a few special cases
(e.g. an ellipsoidal particle), finding configurations
$\hat\mathbf{m}(\mathbf{r})$ of lowest energy is a difficult
computational problem.

\begin{figure*}
\includegraphics[width=0.65\columnwidth]{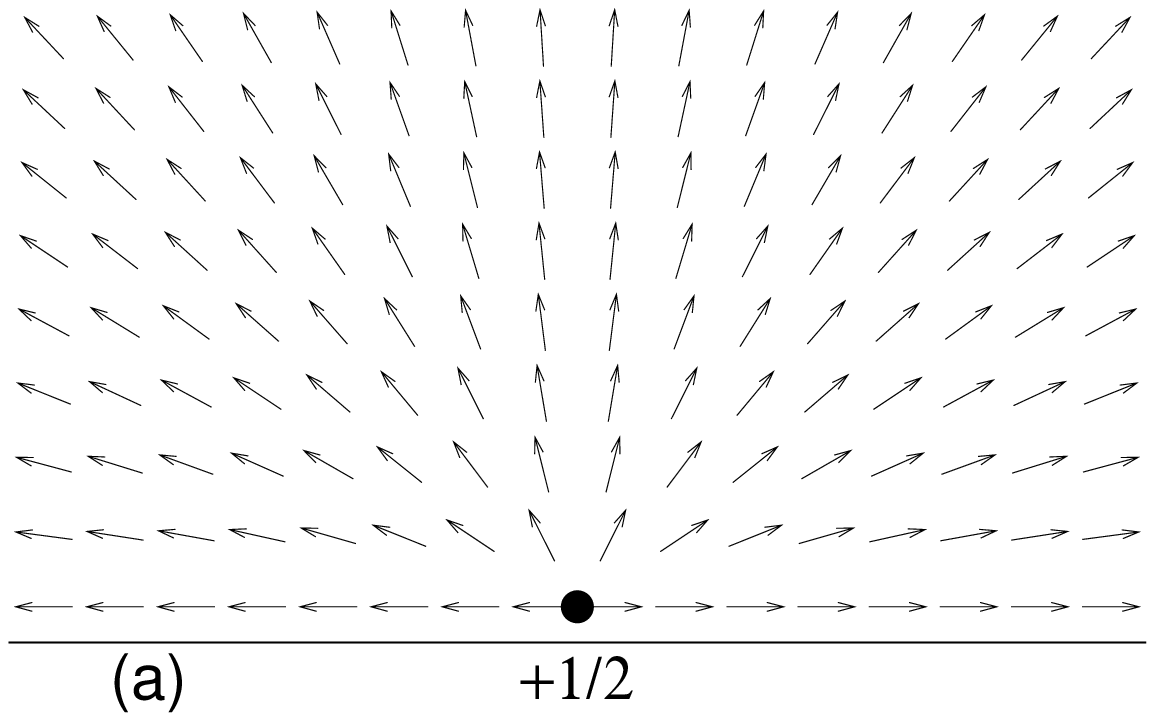}
\includegraphics[width=0.65\columnwidth]{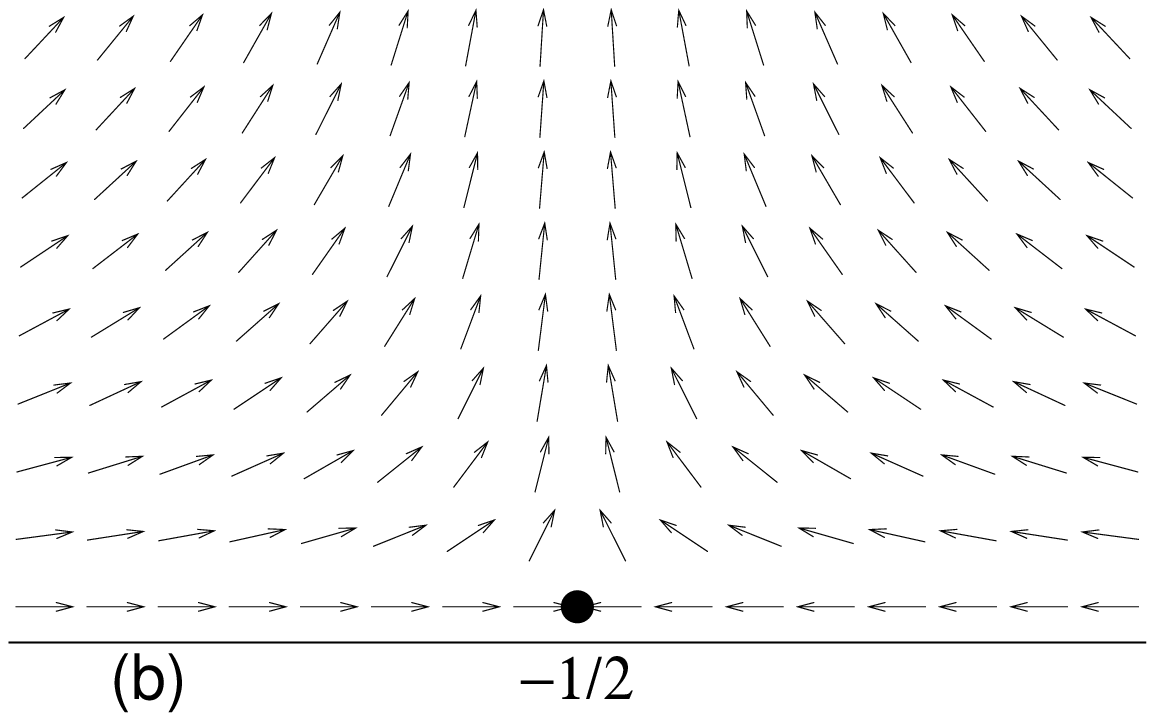}
\includegraphics[width=0.65\columnwidth]{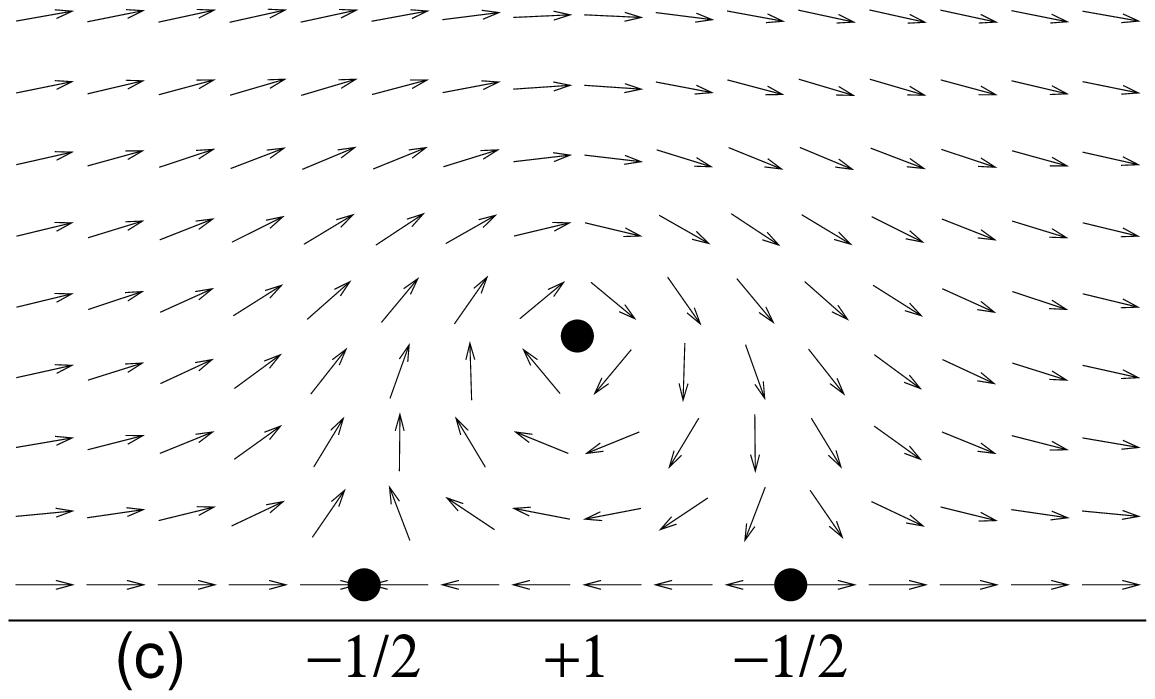}
\includegraphics[width=0.65\columnwidth]{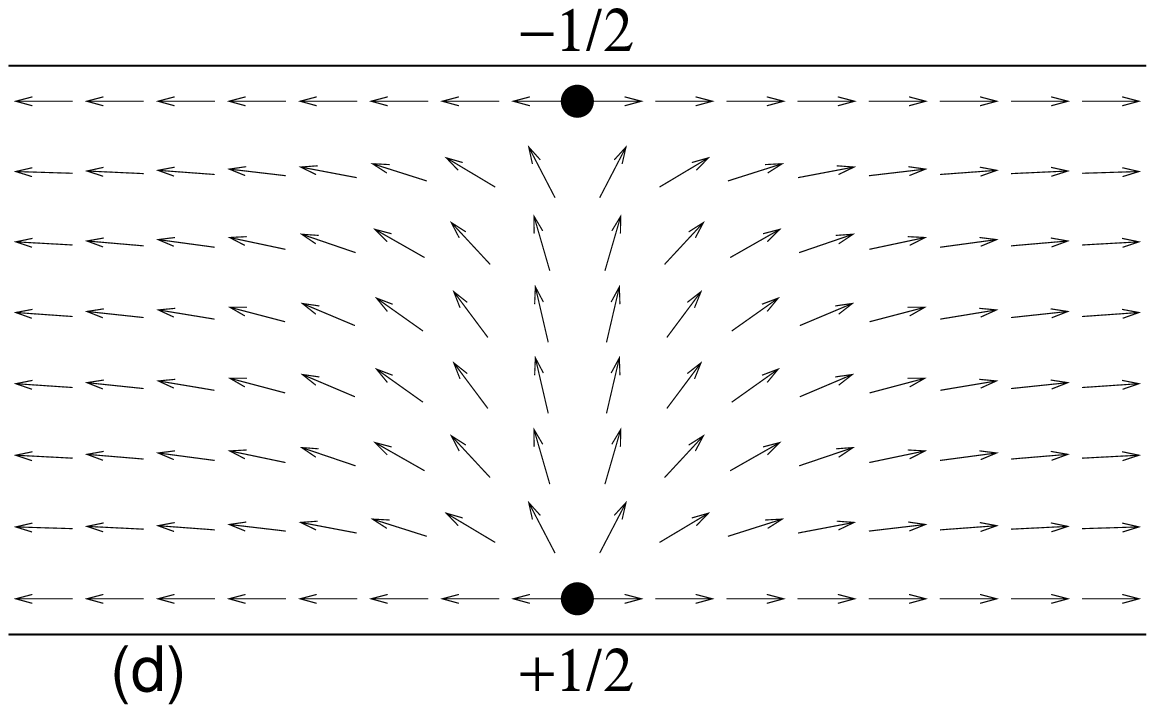}
\includegraphics[width=0.65\columnwidth]{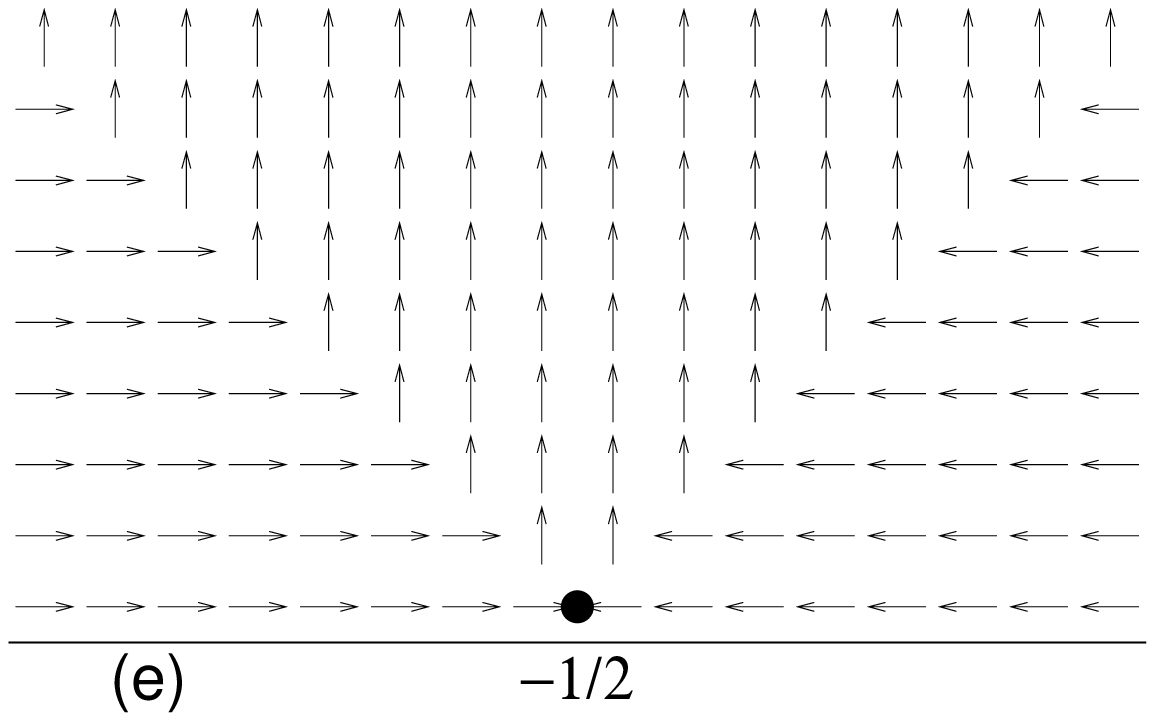}
\caption{(a) and (b) Edge defects with winding numbers $n =
+\frac{1}{2}$ and $-\frac{1}{2}$.  (c) A vortex ($n = +1$) offsets the
influence of two $-\frac{1}{2}$ edge defects.  (d) A domain wall
composed of two edge defects $\pm\frac{1}{2}$.  (e) The $-\frac{1}{2}$
defect in the limit where magnetostatic energy dominates.  }
\label{fig-defects}
\end{figure*}

Analytical treatment is nonetheless possible in a thin-film limit
\cite{Kurzke04,Kohn04} defined for a strip of width $w$ and thickness
$t$ as
\begin{equation}
t \ll w \ll \lambda^2/t \ll w \log{(w/t)}, 
\label{eq-range}
\end{equation}
where $\lambda = \sqrt{A/\mu_0 M^2}$ is a magnetic length scale (5 nm
in permalloy).  Taking this limit yields three simplifications: (a)
magnetization lies in the plane of the film, $\hat\mathbf{m} =
(\cos{\theta}, \sin{\theta}, 0)$; (b) it depends on the in-plane
coordinates $x$ and $y$, but not on $z$; (c) the magnetic energy
becomes a local functional of magnetization \cite{Kurzke04,Kohn04}:
\begin{equation}
E[\hat\mathbf{m}(\mathbf{r})]/At 
= \int_{\Omega} |\nabla \hat\mathbf{m}|^2 \, d^2 r
- (1/\Lambda)\int_{\partial \Omega} 
(\hat\mathbf{m} \cdot \hat{\bm \tau})^2 \, d r.
\label{eq-Kurzke}
\end{equation}
The quantity $E/At$ is a dimensionless energy, $\Omega$ is the
two-dimensional region of the film, $\partial \Omega$ is its line
boundary, $\hat{\bm \tau} = (\cos{\theta_\tau}, \sin{\theta_\tau}, 0)$
is the unit vector parallel to the boundary, and $\Lambda = 4\pi
\lambda^2 / t \log{(w/t)}$ is the effective magnetic length in the
thin-film geometry.  The first---exchange---term in
Eq.~(\ref{eq-Kurzke}) is the familiar XY model \cite{Chaikin} with the
ground states obeying the Laplace equation $\nabla^2\theta = 0$ in the
bulk.  The second term expresses the magnetostatic energy due to
magnetic charges at the film edge and sets the boundary conditions.

Things are particularly simple in the limit $\Lambda/w \to 0$, when
magnetization at the edge is forced to be parallel to the boundary,
$\hat\mathbf{m} = \pm \hat{\bm \tau}$.  Solutions in this limit can be
constructed by exploiting the analogy between the XY model and
electrostatics in two dimensions \cite{Chaikin}, whereby the angle
gradients are associated with components of the electric field, $(E_x,
E_y) = (\partial_y \theta, -\partial_x \theta)$.  Topological
defects---vortices with winding numbers $n = \pm 1$---become positive
and negative point charges of unit strength.  In the presence of an
edge, solutions are obtained by the method of images.  For instance, a
single vortex (winding number $n = \pm 1$) located at $(X,Y)$ in the
infinite semiplane $y>0$ is
\begin{equation}
\theta(x,y) = \pm\arctan{\left(\frac{y-Y}{x-X}\right)}
\pm \arctan{\left(\frac{y+Y}{x-X}\right)}.
\label{eq-vortex}
\end{equation}
The image ``charge'' at $(X,-Y)$ has the same sign as the original.
Therefore, a vortex is repelled by the boundary.

Another class of solutions [Fig.~\ref{fig-defects}(a) and (b)] has a
singularity at the edge \cite{Kurzke04}:
\begin{equation}
\theta(x,y) = \pm\arctan{\left(\frac{y}{x-X}\right)}.
\label{eq-half}
\end{equation}
In the spirit of Eq.~(\ref{eq-vortex}), this solution can be viewed as
a {\em halfvortex} whose winding number $\pm \frac{1}{2}$ is doubled
by the reflection in the edge.  Confinement to the edge occurs because
vortices with noninteger winding numbers cannot exist in the bulk.  An
attempt to peel a halfvortex away from the edge will make $\theta(x,y)$
a multivalued function:
\begin{equation}
\theta(x,y) = \pm\frac{1}{2}\arctan{\left(\frac{y-Y}{x-X}\right)}
\pm \frac{1}{2}\arctan{\left(\frac{y+Y}{x-X}\right)}.
\label{eq-half-bulk}
\end{equation}
The multivaluedness means that the magnetization will be discontinuous
across a line defect extending from the core of the halfvortex to the
film boundary and thus providing a potential $V \propto Y$ confining the
halfvortex to the edge.

The assignment of fractional winding numbers to edge defects can be
justified in a more direct way: a superposition of two edge defects
and one vortex ($n = +1$) has zero total defect strength
[Fig.~\ref{fig-defects}(c)].  Hence $n=-\frac{1}{2}$ for each of the
edge defects.  A model-independent, purely geometrical justification
will be given below.

In a strip $|y| < w/2$, a domain wall interpolating between the
$\theta = 0$ and $\pi$ ground states can be constructed out of two
edge defects with opposite winding numbers $\pm\frac{1}{2}$ and
$\mp\frac{1}{2}$ [Fig.~\ref{fig-defects}(d)]:
\begin{equation}
\tan{\theta(x,y)} = \pm \frac{\cos{(\pi y/w)}}{\sinh{(\pi (x-X)/w)}}.
\label{eq-wall}
\end{equation}
It is clear from the electrostatic analogy that the two defects
experience an attractive ``Coulomb'' force (not strong enough to
overcome the edge confinement).  This force holds the composite domain
wall together.  Because the XY model itself contains no length scale,
the extent of the domain wall along the strip is set by the width of
the strip $w$.  This is precisely the kind of a domain wall observed
in numerical simulations \cite{McMichael97}.

The above solutions can be easily adopted to the cases when the ratio
$\Lambda/w$ is small but finite.  The singular cores of the
halfvortices reside outside the film, the distance $\Lambda$ away from
the edge \cite{Kurzke04}.

\begin{figure*}
\includegraphics[width=0.49\columnwidth]{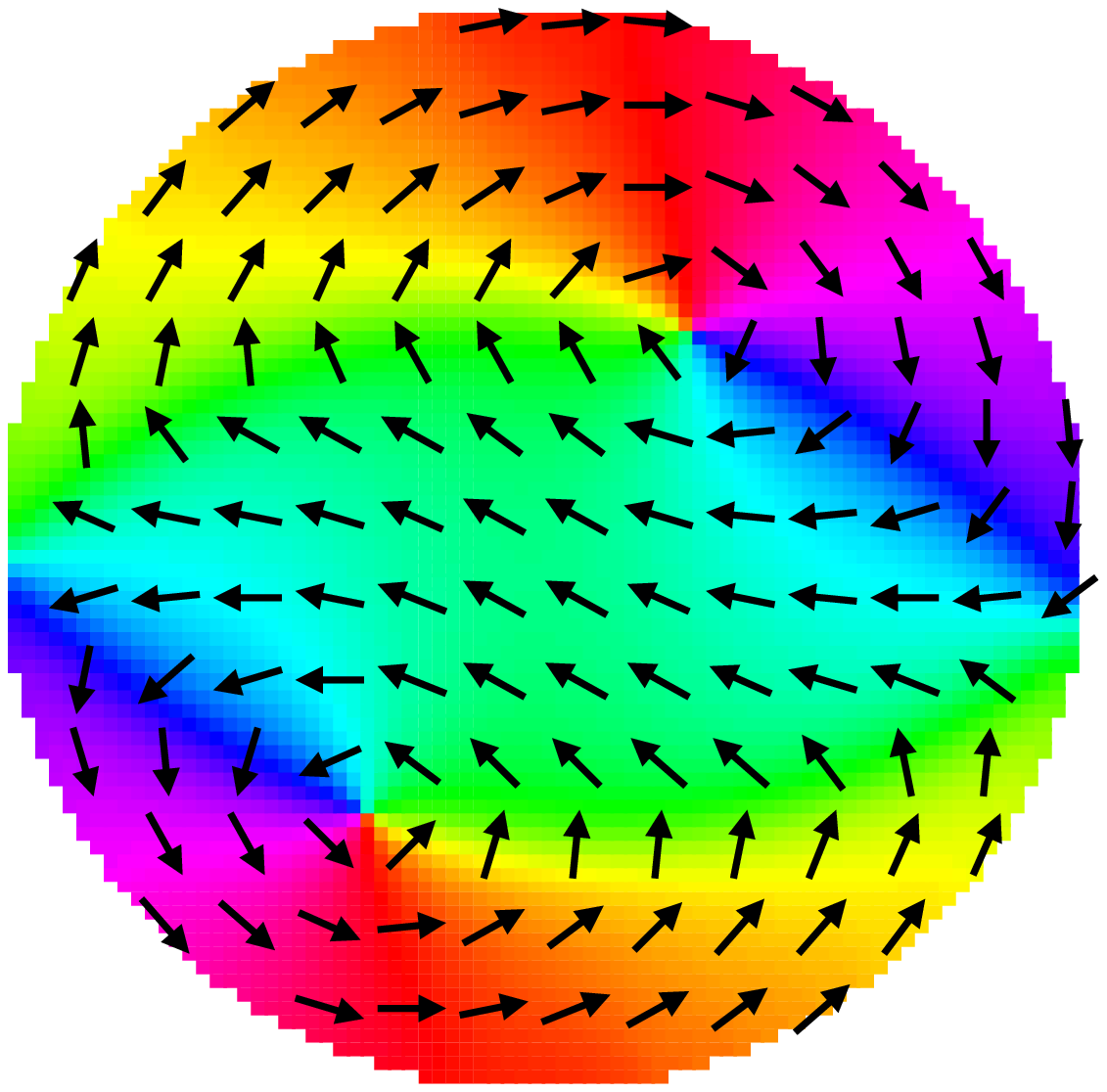}
\includegraphics[width=0.49\columnwidth]{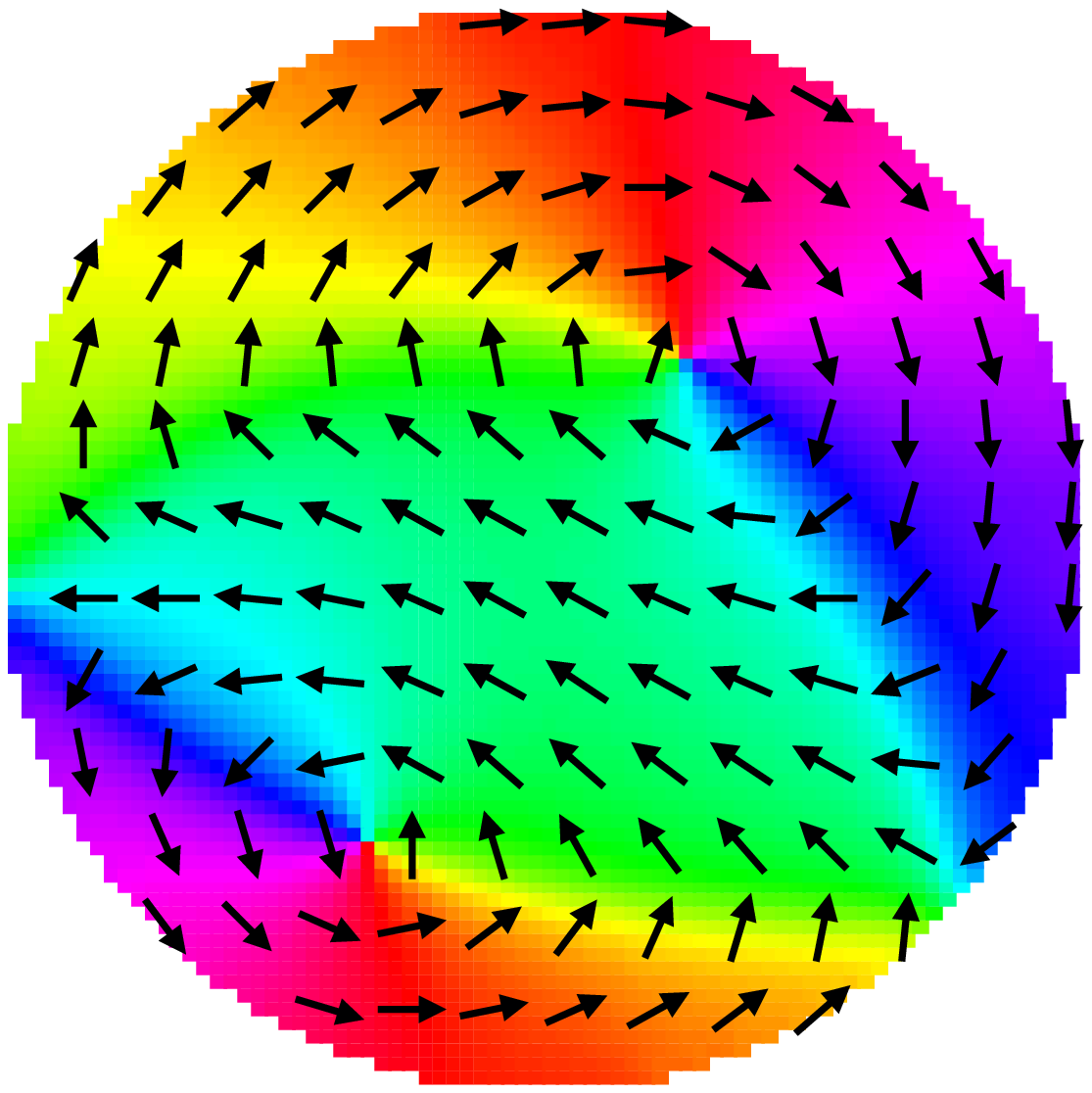}
\includegraphics[width=0.49\columnwidth]{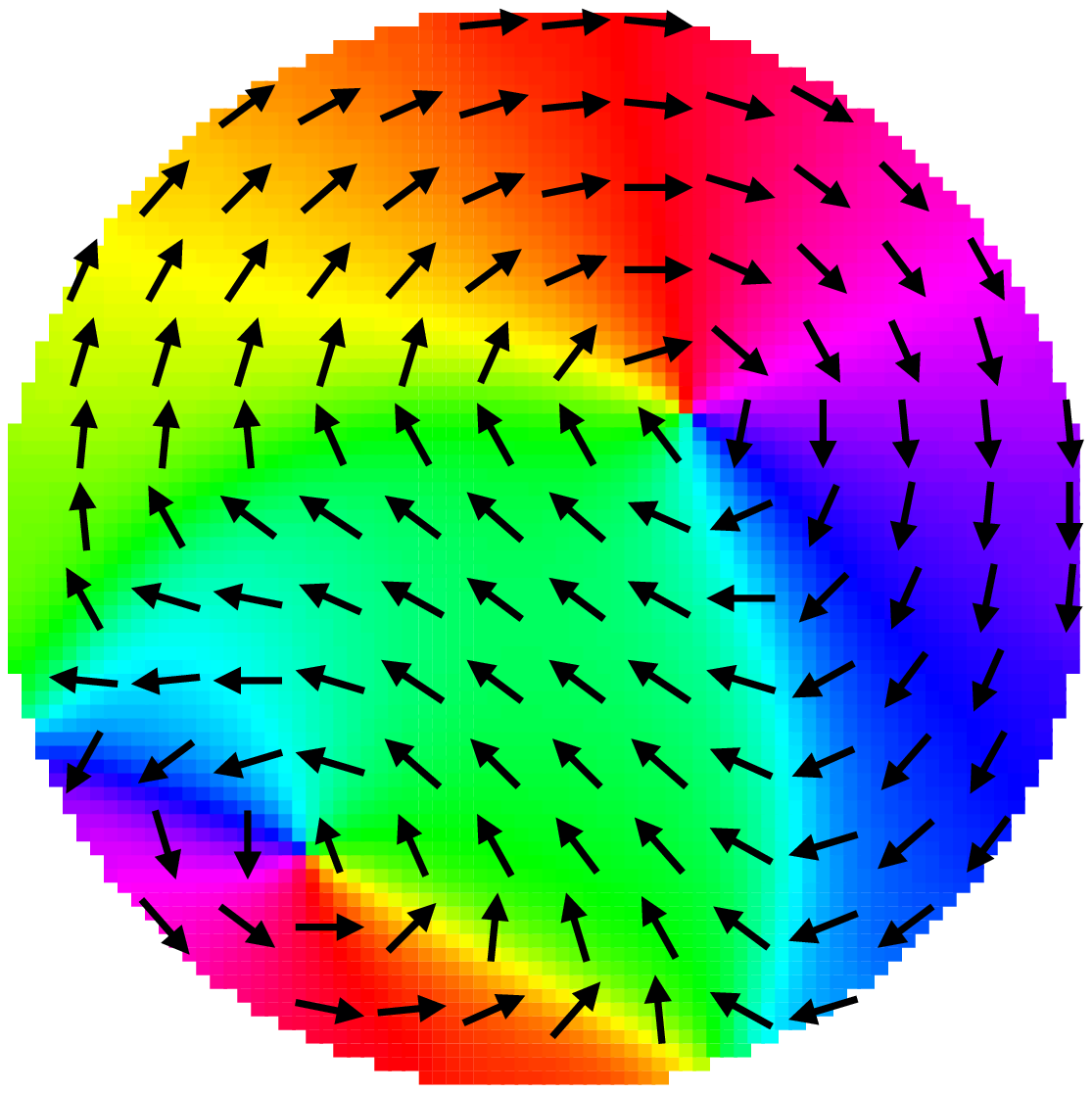}
\includegraphics[width=0.49\columnwidth]{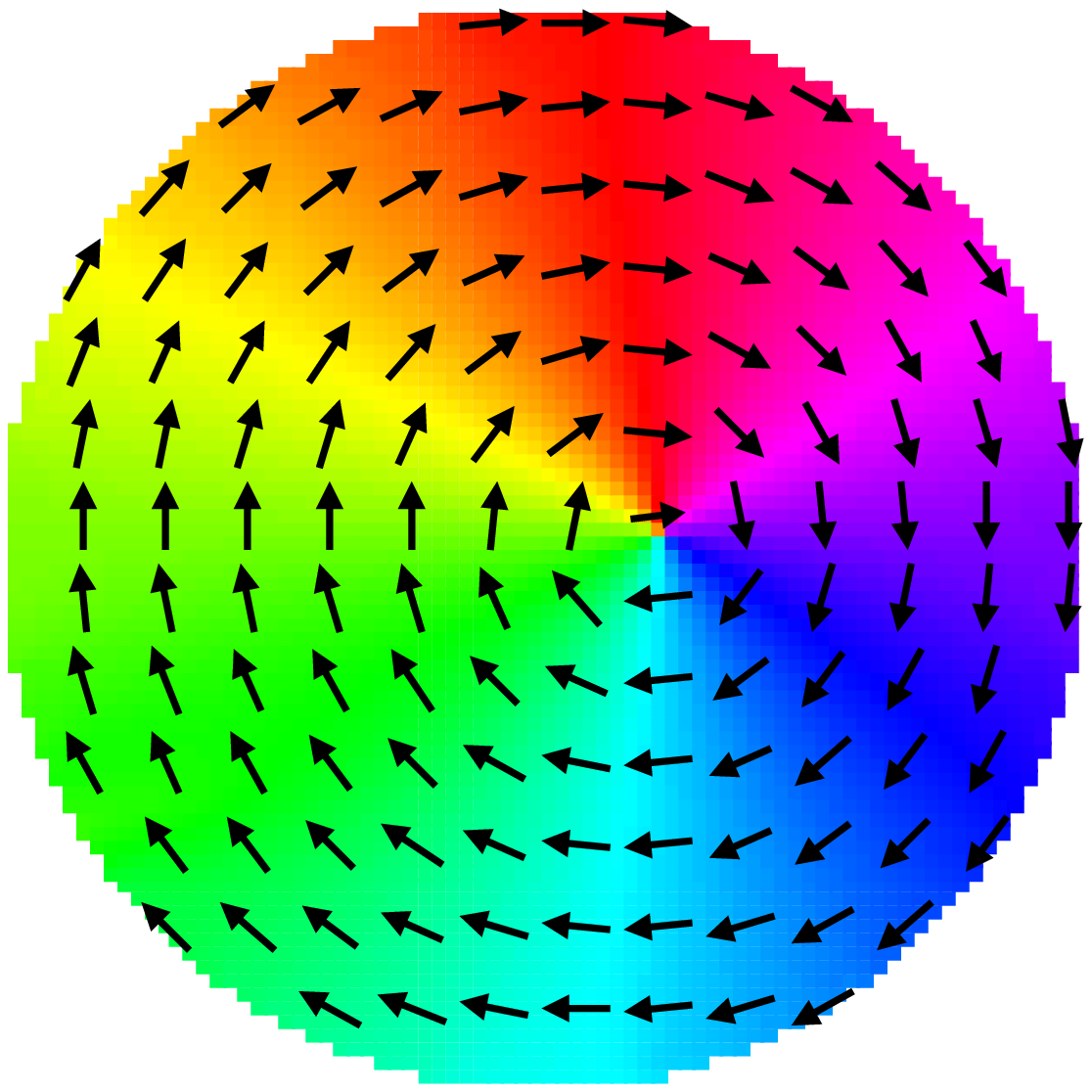}
\caption{Numerical simulation of magnetization dynamics in a permalloy
disk.  Color encodes the direction of the field $\hat\mathbf{m}$ (also
shown by arrows).  Convergence of different colors at a point signals
the presence of a topological defect.  In the first three panels the
disk contains two $+1$ vortices and two $-\frac{1}{2}$ edge defects.
After the edge defects and a vortex mutually annihilate (last panel),
a single vortex remains.  }
\label{fig-oommf}
\end{figure*}

The topological defects identified above survive well outside the
range of applicability of the model (\ref{eq-Kurzke}).
Fig.~\ref{fig-oommf} shows simulated dynamics in a permalloy disk of
the diameter $w = 400$ nm and thickness $t = 20$ nm.  (We used the
numerical package OOMMF \cite{oommf}.) Inequality (\ref{eq-range})
fails badly, yet the bulk and edge defects are easily recognisable and
the winding numbers are clearly conserved.

In wider and thicker strips ($wt \gtrsim \lambda^2$), the
magnetostatic energy breaks the symmetry between the defects with
positive and negative winding numbers inherent to the XY model.  In
particular, the $+\frac{1}{2}$ edge defects have a higher
magnetostatic energy than their $-\frac{1}{2}$ counterparts.  Evidence
for this can be seen in numerical simulations \cite{McMichael97}
showing a more substantial broadening of the cores in $n=+\frac{1}{2}$
defects.

In the magnetostatic regime $wt \gg \lambda^2$, the defects can again
be constructed explicitly \cite{Tch05}.  The magnetostatic energy is
minimized if the density of magnetic charges vanishes in the bulk,
${\bm \nabla} \cdot \hat\mathbf{m} = 0$, and on the surface,
$\hat\mathbf{m} = \pm \hat{\bm \tau}$.  It can be checked that the
$+1$ vortex in the bulk retains its shape [Fig.~\ref{fig-defects}(c)].
The $-\frac{1}{2}$ edge defect looks like two domain walls emanating
from a point at the edge in orthogonal directions
[Fig.~\ref{fig-defects}(e)].  The $-1$ vortex is a similar
intersection of four domain walls.  The V-shaped $-\frac{1}{2}$ edge
defects can be seen in numerical simulations \cite{McMichael97}.
Domain walls in this limit are made of two $-\frac{1}{2}$ defects and
a $+1$ vortex between them \cite{Tch05}.  This type of the domain wall
is favored when $w t \gtrsim C \lambda^2$, where $C \approx 130$ is a
rather large numerical parameter \cite{McMichael97} weakly dependent
on the thickness $t$.

Although different in shape, the $-\frac{1}{2}$ edge defects in the
exchange [Fig.~\ref{fig-defects}(b)] and magnetostatic
[Fig.~\ref{fig-defects}(e)] limits have identical topological
properties.  Generally, as long as magnetization tends to align itself
with the boundary, the edge defects are stable.  They are manifested
as kinks in magnetization $\hat\mathbf{m}$: the latter rotates along 
the edge between the local tangential directions $+\hat{\bm \tau}$ and
$-\hat{\bm \tau}$.  The winding number of a kink is defined as the
line integral along the boundary:
\begin{equation}
n = - \frac{1}{2\pi} \int_{\partial \Omega} {\bm \nabla}(\theta -
\theta_\tau) \cdot d \mathbf{r} = \pm \frac{1}{2}
\label{eq-n-e}
\end{equation}
for a single edge defect.  Because the boundary $\partial \Omega$ is a
closed line, it always contains an even number of kinks.  The sum of
the winding numbers of edge defects is thus an integer related to the
topological charge of bulk defects. In a simply connected (no holes)
region $\Omega$,
\begin{widetext}
\begin{equation}
\sum_{i}^{\text{edge}} n_i = 1 - \frac{1}{2\pi}
\oint_{\partial \Omega} {\bm \nabla}\theta \cdot d \mathbf{r}
= 1- \frac{1}{2\pi}
\int_{\Omega} (\partial_x \partial_y \theta - \partial_y \partial_x \theta)
\, d^2r = 1 - \sum_{i}^{\text{bulk}} n_i,
\label{eq-n}
\end{equation}
\end{widetext}
with integer $n_i$ for bulk defects.  In a film with $g$ holes,
\begin{equation}
\sum_{i} n_i = 1-g.
\label{eq-conserv}
\end{equation}
The winding numbers are $\pm 1$ for bulk defects and $\pm \frac{1}{2}$
for edge defects.  The half-integer value of the topological charge is
directly related to the presence of a kink in magnetization at the
edge \cite{Jackiw76,Goldstone81,Volovik00}.  The halfvortices are
analogs of the boojums thought to exist at an interface between the A
and B phases of superfluid $^3$He \cite{Misirpashaev91,Volovik}.

\begin{figure}
\includegraphics[width=0.65\columnwidth]{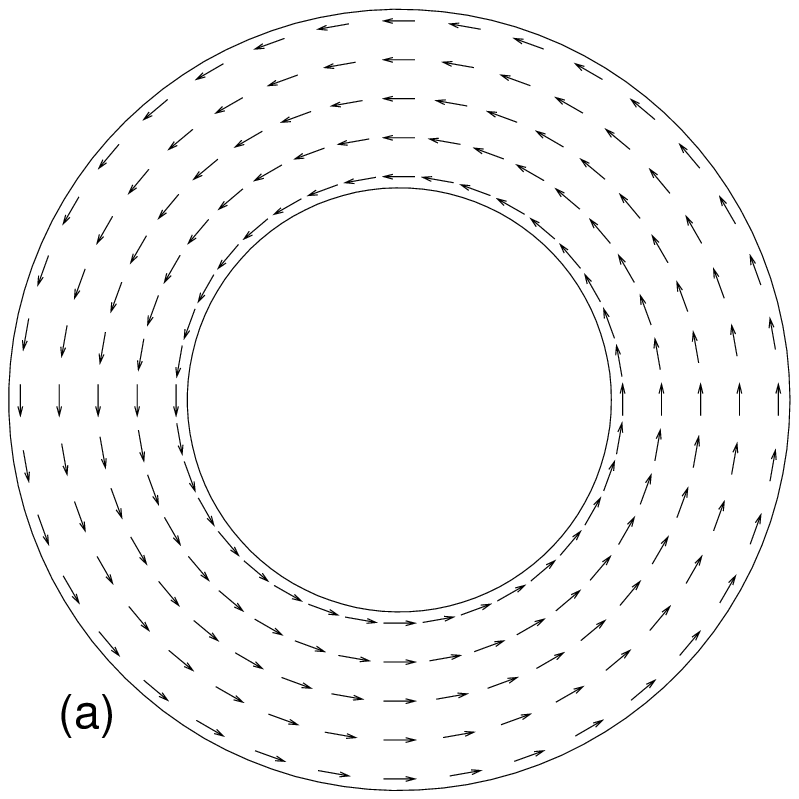}
\includegraphics[width=0.65\columnwidth]{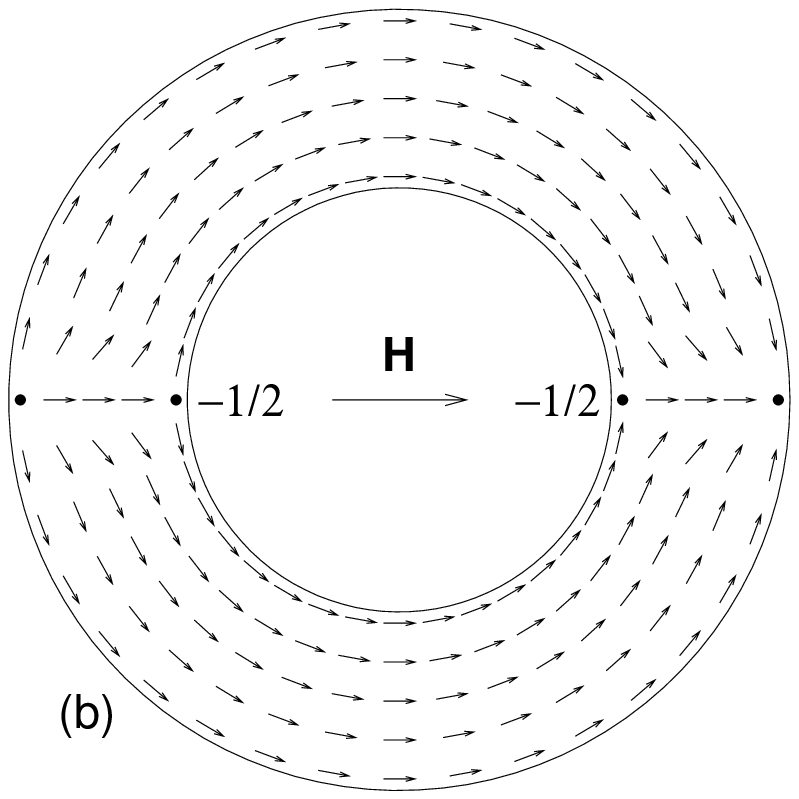}
\caption{Stable and metastable states of a magnetic nanoring in the
thin-film limit. (a) One of the two ground states without topological
defects.  (b) State with a remnant magnetization contains two
composite domain walls with $+\frac{1}{2}$ defects at the outer edge
and $-\frac{1}{2}$ defects at the inner edge.}
\label{fig-ring}
\end{figure}

Conservation of topological charge (\ref{eq-conserv}) has important
implications for the dynamics of magnetization in nanomagnets.  For
example, in rings of certain sizes the ground state contains no
topological defects and has zero magnetic dipole moment
[Fig.~\ref{fig-ring}(a)].  The ring can be magnetized by applying a
strong in-plane magnetic field.  Switching off the field leaves the
ring in a metastable state with remnant magnetization containing two
composite domain walls [Fig.~\ref{fig-ring}(b)].  By applying the
magnetic field in the opposite direction, the walls can be set in
motion on a collision course and may annihilate, leaving the magnet in
a ground state \cite{Klaeui03R}. However, direct annihilation of the
two domain walls is impossible: both walls have the $-\frac{1}{2}$
defects at the inner edge of the ring and two such defects cannot
annihilate on their own.  Accordingly, in thin and narrow rings the
two domain walls do not annihilate but instead form a ``$360^\circ$
domain wall'' \cite{Castano03}.  In thicker and wider rings, the
annihilation does occur and the ring returns to a ground state.  To
facilitate the annihilation, the edge defects in one of the domain
walls must trade places.  Because the edge defects themselves cannot
move into the bulk, they alter the signs by exchanging a vortex.  The
$+\frac{1}{2}$ defect emits a $+1$ vortex into the bulk and converts
into a $-\frac{1}{2}$ defect \cite{Vavassori04}.  The emitted vortex
travels to the inner edge where it fuses with the $-\frac{1}{2}$
defect into a $+\frac{1}{2}$ defect.  Now the defects at both edges
can annihilate directly.  In thin and narrow rings this does not
happen because vortex emission is forbidden
{\em energetically} \cite{McMichael97}.  Transient states with vortices in the
bulk of the ring have been observed in simulations \cite{Klaeui03R}.
However, their role in catalysing the annihilation of domain walls has
not been appreciated.

We have uncovered a simple structure underlying intricate magnetic
patterns in soft nanomagnets in a planar geometry.  The patterns are
formed by a few highly stable topological defects: ordinary vortices
in the bulk and halfvortices confined to the edge.  The importance of
edge defects in nanomagnets has been overlooked.  Here we have
demonstrated that the ``transverse'' domain walls found in thin and
narrow strips \cite{McMichael97} are made of two such edge defects
wiht opposite winding numbers.  We will describe elsewhere
\cite{Tch05} a simple model of the ``vortex'' domain walls
\cite{McMichael97} made of two $-\frac{1}{2}$ edge defects and a $+1$
vortex.  It appears that the general approach taken in this paper may
provide a basic model of complex magnetization dynamics in nanomagnets
by reducing it to the creation, propagation, and annihilation of a few
topological defects.

{\bf Acknowledgments.}  We thank C.-L. Chien, P. Fendley,
R. L. Leheny, I. Tchernyshyov, and F. Q. Zhu for helpful discussions.
The work was supported in part by the NSF Grant No. DMR05-20491.

\end{document}